\documentclass[twocolumn]{aastex63}

\def    \apjl  		{\rm {ApJL}}

\def    \apjl  		{\rm {ApJL}}

\usepackage{savesym}
\savesymbol{tablenum}
\usepackage{siunitx}
\restoresymbol{SIX}{tablenum}
\usepackage{csquotes}

\accepted{by ApJ on October 9, 2021}
\shorttitle{Spatially varying temperatures in IC\,63 PDR}
\shortauthors{Soam et al.}
\graphicspath{{./}{figures/}}

\begin{document}
\title{Spatial variation in temperature and density in the IC\,63 PDR from $\rm H_{2}$ Spectroscopy}

\correspondingauthor{Archana Soam}
\email{asoam@usra.edu}

\author[0000-0002-6386-2906]{Archana Soam}
\affiliation{SOFIA Science Center, Universities Space Research Association, NASA Ames Research Center, M.S. N232-12, Moffett Field, CA 94035, USA}
\author[0000-0001-6717-0686]{B-G Andersson}
\affiliation{SOFIA Science Center, Universities Space Research Association, NASA Ames Research Center, M.S. N232-12, Moffett Field, CA 94035, USA}
\author[0000-0001-5996-3600]{Janik Karoly}
\affiliation{Jeremiah Horrocks Institute, University of Central Lancashire, Preston PR1 2HE, UK}
\affiliation{SOFIA Science Center, Universities Space Research Association, NASA Ames Research Center, M.S. N232-12, Moffett Field, CA 94035, USA}
\author{Curtis DeWitt}
\affiliation{SOFIA Science Center, Universities Space Research Association, NASA Ames Research Center, M.S. N232-12, Moffett Field, CA 94035, USA}
\author{Matthew Richter}
\affiliation{Department of Physics and Astronomy, University of California, Davis, CA, USA}



\begin{abstract}

We have measured the gas temperature in the IC\,63 photodissociation region (PDR) using the S(1) and S(5) pure rotation lines of molecular hydrogen with SOFIA/EXES. We divide the PDR into three regions for analysis based on the illumination from $\gamma$ Cas: \enquote{sunny}, \enquote{ridge}, and \enquote{shady}. Constructing rotation diagrams for the different regions, we obtain temperatures of T$_{ex}$=$562^{+52}_{-43}$\,K towards the \enquote{ridge} and T$_{ex}$=$495^{+28}_{-25}$\,K in the \enquote{shady} side. The H$_2$ emission was not detected on the \enquote{sunny} side of the ridge, likely due to the photo-dissociation of H$_2$ in this gas. Our temperature values are lower than the value of T$_{ex}$=685$\pm$68\,K using the S(1), S(3), and S(5) pure rotation lines,  derived by \citet{2009MNRAS.400..622T} using lower spatial-resolution ISO-SWS data at a different location of the IC\,63 PDR.  This difference indicates that the PDR is inhomogeneous and illustrates the need for high-resolution mapping of such regions to fully understand their physics. The detection of a temperature gradient correlated with the extinction into the cloud, points to the ability of using H$_2$ pure rotational line spectroscopy to map the gas temperature on small scales. We used a PDR model to estimate the FUV radiation and corresponding gas densities in IC\,63. Our results shows the capability of SOFIA/EXES to resolve and provide detailed information on the temperature in such regions. 
\end{abstract}



\section{Introduction} \label{sec:intro}

Molecular hydrogen is readily detectable in photo-dissociation regions (PDRs; \citealt{Nadeau1991, 1996A&A...315L.281T, 2004ARA&A..42..119V}). Hydrogen transitions from $\rm H_{2}$, via H\ I to H\ II in PDRs from the exposure to far-ultra violet (FUV) radiation from O/B stars and the interstellar radiation field. Through photochemical reactions and photo-electric emission, both the chemistry and temperature balance of the PDRs is controlled by the FUV radiation \citep{2007A&A...467..187R}. It is therefore important to characterise the over-all as well as small scale structure of these regions.

In the regions directly affected by FUV radiation, the thermal balance is related to the radiative transfer of the UV photons. The FUV radiation is also responsible for the kinematics and chemical changes of these regions \citep{draineBertoldi1999}. The ejection of photo-electrons from the dust grains is the the process accountable for heating of the region \citep{bakesTielens1994, weingartner2001}. In addition to this process, the collisional de-excitation of $\rm H_{2}$ molecules initially excited by UV photons is also a mechanism contributing to the heating \citep{sternberg1989}. Towards the higher density regimes, gas-grain collisions may also heat the gas. On the other hand, the fine-structure lines of neutral atoms or of singly ionized species provide the cooling of the outer layers and CO rotational lines cool the predominantly molecular inner regions \citep{allers2005}.

The observations of rotational and vibrational lines of molecular hydrogen is a useful tool to diagnose the physical properties of the PDRs. As investigated by \citet{allers2005} towards Orion bar, the low critical density of the ground-state rotational transitions makes ratios of mid-IR lines good probes of the temperature in the layers where they arise. In PDRs with $\rm n(H) \sim 10^4-10^5\,cm^{-3}$, the lower rotational levels of $\rm H_{2}$ (i.e. J=2-0, J=3-1) are maintained in thermal equilibrium by collisions \citep{2005SSRv..119...71H}. The critical density of $\rm n(H) > 10^5\,cm^{-3}$ is needed for higher transitions (i.e. J=5-3, J=7-5). The populations of these levels can be a good tracer of gas temperature.

We investigated the pure rotational molecular hydrogen line emission in the PDR IC\,63. This is a small, but well-studied (at d$\approx$200\,pc; \citealt{karr2005}), PDR irradiated by the (FUV) light from $\gamma$ Cas, a B0.5\ IV type star located $\sim$1.3\,pc projected distance from the cloud. This is the nearest and well explored H\,II system to investigate the PDR properties. The system contains nebaule IC\,59 and IC\,63 irradiated by $\gamma$ Cas. The projected direction of FUV radiation from $\gamma$ Cas on IC\,63 is from southwest to northeast.


\begin{table*}
	\centering
	\caption{Details of SOFIA/EXES observations of IC\,63}
	\scriptsize
	\begin{tabular}{ccccccc}\hline
UT DATE	 & Line transition   & $\rm \lambda_{central}$  & 	Central coordinate	      &	       $\rm VPA^{\dagger}$ range      & Doppler offset &	Int. time  \\ 
&   &($\mu$m)       &           (RA$_{J2000}$, Dec$_{J2000}$)  & ($^{\circ}$)    & ($\rm km\,s^{-1}$) & (s)  \\  \hline
2018-10-19  &S(1) 3-1	&	17.03	&	00:58:59.8600,$+$60:53:39.900   &	       97.4377-26.1064 &	-12.14	  &  	2704  \\
2018-10-20  &S(5) 7-5	&	6.91	&	00:59:00.2700,$+$60:53:45.300   &	       93.1134-24.7629 &	-11.81	  &	5532   \\
2018-10-24  &S(5) 7-5	&	6.91	&	00:58:59.6000,$+$60:53:45.300   &	       83.2769-24.613  &	-10.52	  &  	4992 \\
2018-10-25  &S(1) 3-1	&	17.03	&	00:58:59.6000,$+$60:53:45.300   &	       59.3702-48.2404 &	-10.17	  &  	896  \\ \hline
	\end{tabular}
	\label{tab:observations}
	
$\dagger$ Vertical Position Angle (VPA), defined as Degrees East of North. 
\end{table*}

The goal of SOFIA proposal 07\_0040 (PI: Soam, A.) was to quantify the difference in gas temperature in gas directly exposed to the light from $\gamma$ Cas, and gas in the shade of a molecular clump within the PDR. Here, we present results of observations of the S(1) and S(5) lines of H$_2$. We compare our SOFIA/EXES (Stratospheric Observatory for Infrared Astronomy/Echelon-Cross-Echelle Spectrograph) \citep{2018JAI.....740013R} findings with the results of \citet{2009MNRAS.400..622T} toward a different location of the IC\,63 PDR.  As is often the case in the ISM \citep{1998ASSL..236.....E}, \citet{2009MNRAS.400..622T}, using observations of S(0), S(1), S(3), and S(5) from the Short Wavelength Spectrometer (SWS) on board the Infrared Space Observatory (ISO), found a two-temperature structure in the H$_2$ excitation. The two lowest rotation transitions found a low temperature, typical of the cold neutral medium \citep{1977ApJ...218..148M}, in their case T$_{ex}$=106$\pm$11\,K, while their three highest transitions, yield a significantly higher temperature of T$_{ex}$=685$\pm$68K.

Such high(er) excitation temperatures in the high J lines of H$_2$ are commonly seen in FUV absorption spectroscopy and has been variously attributed to the effects of UV pumping via the Lyman and Werner band excitations \citep[e.g.][]{1975ApJ...197..581J} and to collisional heating, including through shocks.  

\citet{andrews2018} investigated the physical conditions of UV-illuminated surface of IC\,63. Using archival \textit{Spitzer} IRS data of molecular hydrogen in IC\,63, they constructed the excitation diagram to find gas temperature in IC\,63 PDR. Assuming that lines are optically thin, adopting an ortho-to-para ratio of three, and using energies and Einstein coefficients from \citet{rosenthal2000}, they found two temperature components in IC\,63 PDR.

The observed correlation between the column densities of the CH$^+$ ion (whose formation, via the reaction of C$^+$ with H$_2$, includes a $\sim$4600\,K activation barrier; \citealt{1991A&AS...87..585M}) and H$_2$ J=3 and J=5 \citep[e.g.][]{1986ApJ...303..401L} indicates that collisional excitation does contribute to the high-J population of H$_2$.  


\begin{figure*}
	\centering
    \includegraphics[scale=0.5,angle=0]{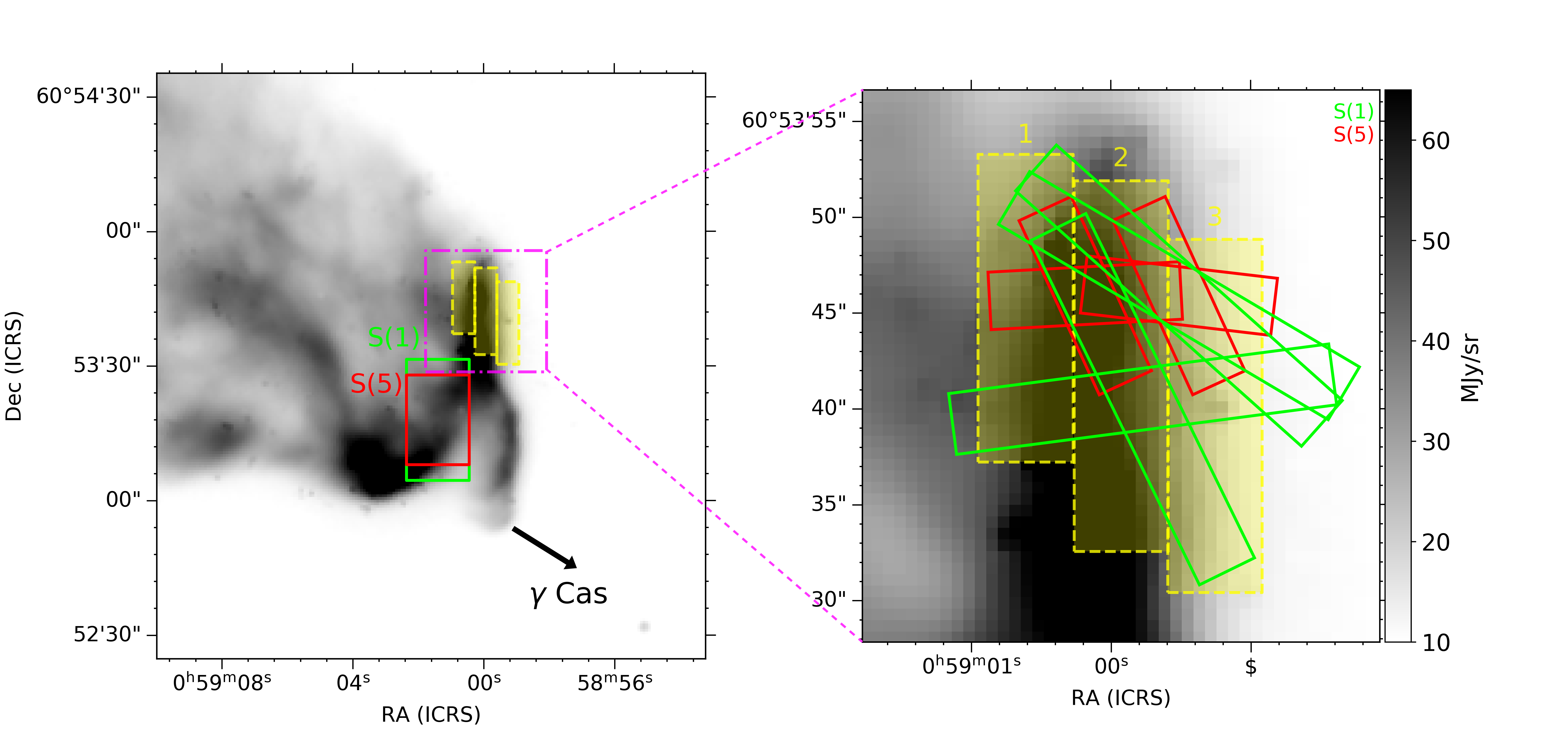}
	\caption{Left panel shows the location of the ISO/SWS slits and the three EXES regions over an 8.0$\micron$ \textit{Spitzer} IRAC image (taken from \textit{Spitzer} archive). The larger solid green rectangle is the ISO/SWS S(1) (17.0$\micron$) aperture and the red solid rectangle is the S(5) (6.9$\micron$) aperture \citep{2009MNRAS.400..622T}. The purple dot-dash line shows the zoomed in image in the right panel and the yellow rectangles show the three regions observed by SOFIA/EXES. The direction to $\gamma$ Cas is shown as well. The right panel shows the zoomed-in view of the three extraction regions and the EXES slit positions over the 8.0$\micron$ IRAC image. The SOFIA/EXES S(1) aperture is the larger solid green rectangles and the S(5) aperture is the smaller solid red rectangles. The slits swept clockwise during observations. We have defined region 1 as \enquote{shady}, region 2 as \enquote{ridge} and region 3 as \enquote{sunny} and they are shown as translucent yellow rectangles.}
	\label{fig:slit}
\end{figure*}

In PDRs, FUV pumping, photo-dissociation, and photo-electric emission are critical processes. On the PDR outer boundary, hydrogen exists mainly in atomic form because most of the molecular hydrogen is photo-dissociated by the high energy UV photons \citep{2009MNRAS.400..622T}. As we move into, or behind, high density/extinction regions inside the PDR, the strength of the UV radiation is reduced by dust extinction, which, together with line self-shielding, reduces the photo-dissociation in comparison to $\rm H_{2}$ formation on the dust grain surface and the molecular fraction increases \citep{habart2004,habart2011}. While the cut-off for photo-electric heating is not at quite as short a wavelength as for H$_2$ destruction (1550\AA\ for an assumed work function of 8\,eV; \citealt{2006ApJ...645.1188W}), the extinction at such wavelengths is still very steep. As the shielding increases, we would therefore expect the gas temperature to drop.  In an inhomogeneous medium where significant shading occurs due to opaque clumps, a (variable) two temperature structure would be expected, corresponding to gas experiencing limited or significant FUV extinction.  Given the size and location of the \citet{2009MNRAS.400..622T} ISO apertures (i.e. $\approx 25^{\prime\prime}$ or 0.025\,pc at the distance of IC\,63), it is conceivable that their two-temperature result originate from a mixture of gas at the front and back of the H $\rightarrow$ H$_2$ transition over their apertures (Figure \ref{fig:slit}). Therefore, we set out to investigate the variation of the temperatures (if present) throughout the PDR, using higher spatial resolution observations than those from ISO/SWS. With its narrow slit, EXES on SOFIA has the capability to produce such higher spectral resolution data for these investigations.

The goal of this paper is to investigate gas temperatures in those regions of IC\,63 PDR which are directly illuminated by and in shade of UV radiation from $\gamma$ Cas. Section \ref{sec:obs} presents the details of SOFIA/EXES observations. Section \ref{sec:res} presents our results and analysis performed. Section \ref{sec:diss} shows the detailed discussion of our results and section \ref{sec:con} concludes our findings. 

\section{Observations}\label{sec:obs}
\subsection{Pure rotational lines}
We observed IC\,63 with SOFIA using the EXES instrument in its high-medium mode \citep{2018JAI.....740013R}, selecting the highest operable cross dispersion order to maximize the spatial length of the slit. For the $\rm H_{2}$ S(1) observations, we observed at 587\, $\rm cm^{-1}$ in $\rm 3^{rd}$ order, with a slit length of 19$^{\prime\prime}$ (0.0184\,pc assuming a distance of $\approx$200\,pc). For $\rm H_{2}$ S(5), we observed at 1447\, $\rm cm^{-1}$ in $\rm 8^{th}$ order, with a slit length of 10$^{\prime\prime}$ (0.0097\,pc). All observations
used the 3.2$^{\prime\prime}$ (0.0031\,pc) slit width, which provides R\,$\sim$60000 resolution, or $\rm \sim 5\,km\,s^{-1}$ (measured with laboratory gas cell data prior to flight (EXES PI team, private communication)). 

We observed IC\,63 using off-source nodding.  We selected an off-region of blank sky 30$^{\prime\prime}$ East of the central slit coordinate (Table \ref{tab:observations}) and nodded at intervals of 60\,s in order to subtract telluric emission and thermal background from the on-source integration.

The spatial resolution of the observations is approximately 3.7$^{\prime\prime}$ for both wavelengths, which is taken from the 50\% encircled energies reported in the SOFIA observatory characterization \citep{temi2018}. The telescope was guided using the Focal Plane Imager Plus (FPI+) guider camera on a visually bright star offset from IC\,63. This mode of guiding is astrometrically accurate to ~0.41$^{\prime\prime}$ with each new target acquisition and with a median pointing stability of 0.17$^{\prime\prime}$ following acquisition \citep{temi2018}.

Flux calibration was done by scaling the data to spectral flat fields with the instrument's external blackbody source before each observation. The resulting absolute flux calibration is good to ~25\%. The relative flux calibration between S(1) and S(5) is 12.5\% but any systematic differences in the relative or the absolute flux calibration are repeatable to 2\% (EXES PI team, private communication).

The rotation of field (ROT) for SOFIA observations is determined by the heading of the airplane but can be held fixed for short periods by the telescope.  Because EXES does not have a field rotator, as some SOFIA other instruments, the slit position angle (PA) changed by discrete amounts every 6 minutes during the observations. We executed individual nodded files within that interval, typically
obtaining 4 nod pairs per file and per discrete PA. Table 1 reports
the starting and ending slit position angles, which are also visualized
in Figure \ref{fig:slit}.

Data were reduced using with the SOFIA Redux
pipeline \citep{2015ASPC..495..355C}, including steps for spike removal,
nod-subtraction, flat fielding, flux calibration, order
rectification, and coadding of nodded pairs.
The wavelength scales were calibrated by
matching the sky emission lines within the spectral settings to their
values in the HITRAN database \citep{gordon2017}. The uncertainty in the wavelength
solution is estimated to be $\rm \sim 0.3\,km\,s^{-1}$.

Because of the changing slit angle during observations, we employed a customized procedure for combining the data. We defined three 5$^{\prime\prime}$-width intervals of RA, as shown in Figure \ref{fig:slit}. The RA centers of the regions were 1) 00:59:00.61, 00:58:59.93 and 3) 00:58:59.25. We used the central slit coordinate and position angle of each file to assign and extract the appropriate parts of the 2-D spectrum to regions 1, 2 and 3, weighting each extracted spatial row by the file's integration time. Because there were up to 20 nodded files and different position angles used during each night's observation, we only list the starting and ending position angle in Table 1.

Our observations of the H$_2$ S(1) and S(5) lines with the narrower EXES slit toward IC\,63 are located in a region chosen to provide as narrow a PDR ridge as possible, somewhat north of the region observed by \citet{2009MNRAS.400..622T}. The location of the EXES slits and the \citet{2009MNRAS.400..622T} apertures are shown on \textit{Spitzer} IRAC image of IC\,63 in Figure \ref{fig:slit} . In the left panel of Figure \ref{fig:slit}, the green and red rectangles show the locations of the ISO/SWS slits for S(1) and S(5) observations of \citet{2009MNRAS.400..622T} while the observed regions by EXES are shown with semi-transparent yellow rectangles. We designated these regions, labeled as 1, 2, and 3 in the right panel of Figure \ref{fig:slit}, as the \enquote{shade}, \enquote{ridge}, and \enquote{sunny} sides, respectively, of the PDR. The right panel of the Figure \ref{fig:slit} also shows the EXES slit rotation pattern and actual extraction regions of S(1) and S(5) emission shown with green and red rectangles, respectively.

The simple vertical RA bars we used for regions were well-motivated by the 2.12\micron~ ro-vibrational $\rm H_{2}$ emission and 8.0\micron~ Spitzer/IRAC emission (Figure \ref{fig:slit}). Because of the exploratory nature of the observations, most of the S(1) was acquired in regions that were close but did not strictly overlap the S(5) coverage. But the full integration time/spatial extent of the S(1) line was needed to obtain a useful line detection. This might mean that the temperature gradient reported here could be mimicked by an inopportune brightness gradient between the total S(1) footprint and the S(5) footprint. However, the 2.12\micron~ ro-vibrational emission and Spitzer/IRAC image do not suggest such a gradient.

\subsection{Near-IR imaging of ro-vibrational line}
We also adopted $\rm H_{2}$ 1-0 S(1) line observations of IC\,63 at 2.122$\mu$m from \citet{2013ApJ...775...84A} for our analysis in this work. They acquired this data from the WIRCam instrument \citep{puget2004} at the Canada–France–Hawaii Telescope (CFHT) in queue observing mode during 2008 August 12-–September 14. In these observations, \citet{2013ApJ...775...84A} used Ks filter to estimate the continuum contributions to the nebular emission because the narrow-band continuum filter associated with the H$_{2}$ 1–0 S(1) filter was not installed at the time of these observations. The data was reduced using observatory pipeline. The photometric calibration of the data was achieved using several hundred 2MASS \citep{Skrutskie2006} stars in the observed field. More details of observations, data reduction, and calibration can be seen in the paper by \citet{2013ApJ...775...84A}.

\section{Results and analysis}\label{sec:res}

\begin{figure*}
	\centering
    \includegraphics[scale=0.75,angle=0]{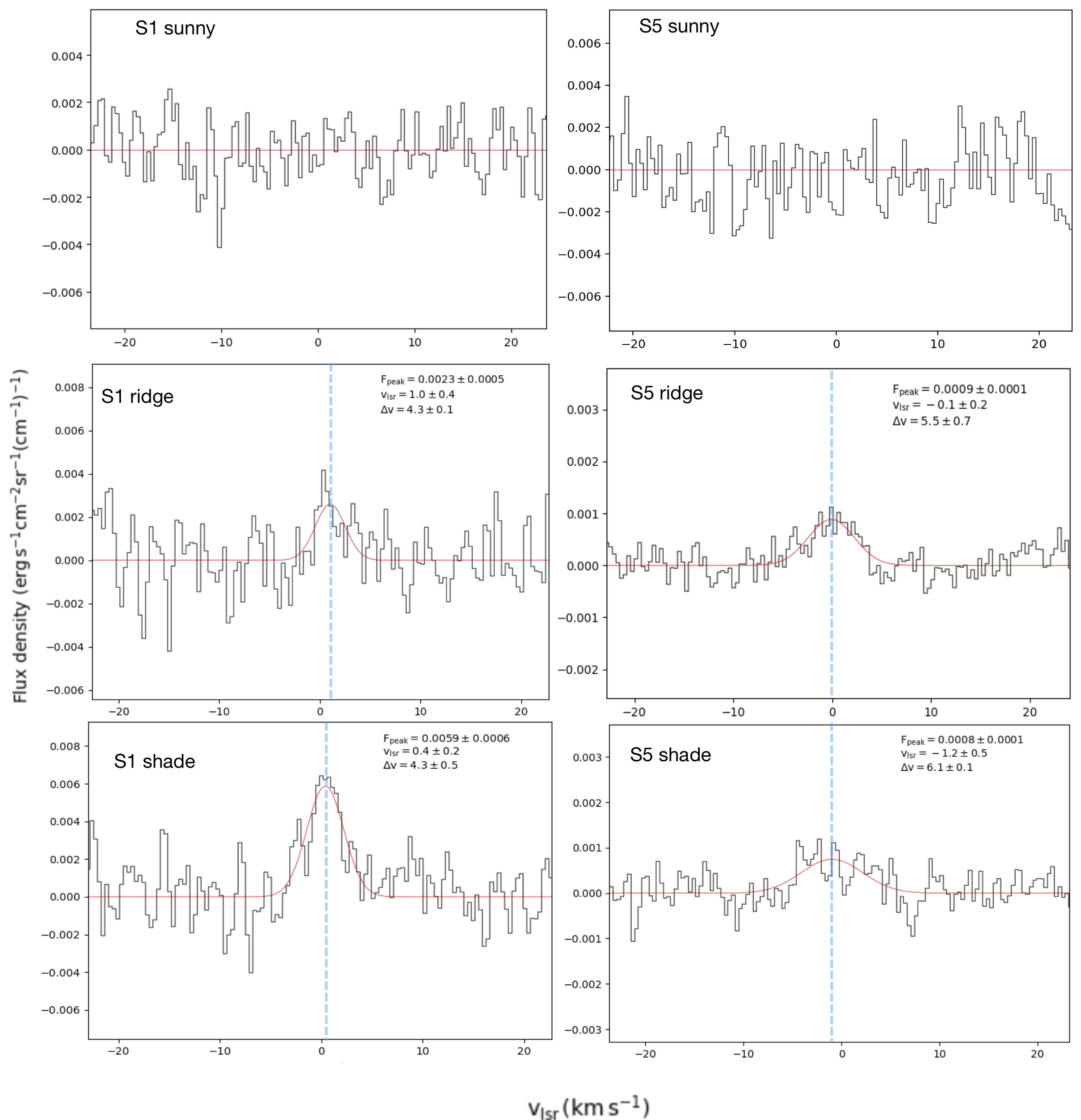}
	\caption{The line profiles of S(1) and S(5) $\rm H_{2}$ emissions in \enquote{ridge}, and \enquote{shady} sides of IC\,63 PDR. No emission is detected in \enquote{sunny} side of the PDR. The dashed blue shows the center of the detected emission and the red solid line shows the Gaussian fit of the emission.}
	\label{fig:lines}
\end{figure*}

\begin{table*}
	\centering
	\caption{Properties of observed $\rm H_{2}$ transitions in IC\,63 PDR.}
	\scriptsize
	\begin{tabular}{ccccccc}\hline
		Line & Reg$^{a}$ & $\lambda$ & $I_{ul}$ & E$_{up}$/k$_{B}$ & A$_{ul}^{b}$ & $\Delta v^{c}$\\
		& & (\micron) & $10^{-5}$ $\rm erg\,\,cm^{-2} s^{-1} sr^{-1}$ & (K) & (s$^{-1}$) & ($\rm km\,s^{-1}$)\\ \hline
		S(1) 3-1 & Shady & 17.035 & 5.3 $\pm$ 0.8 & 1015.12 & 4.8 $\times$ 10$^{-10}$ & 4.3 $\pm$ 0.5\\
		S(1) 3-1 & Ridge & 17.035 & 2.2 $\pm$ 0.8 & 1015.12 & 4.8 $\times$ 10$^{-10}$  & 5.1 $\pm$ 1.5\\
		S(1) 3-1 & Sunny & 17.035 & -- & 1015.12 & 4.8 $\times$ 10$^{-10}$  &  --\\
		S(5) 7-5 & Shady & 6.901 & 2.5 $\pm$ 0.6 & 4586.30 & 5.9 $\times$ 10$^{-8}$ & 6.1 $\pm$ 1.2\\
		S(5) 7-5 & Ridge & 6.901 & 2.6 $\pm$ 0.4 & 4586.30 & 5.9 $\times$ 10$^{-8}$ & 5.5 $\pm$ 0.7 \\
		S(5) 7-5 & Sunny & 6.901 & -- & 4586.30 & 5.9 $\times$ 10$^{-8}$ &  --\\ \hline
	\end{tabular}
	\label{tab:results}
	
	a. See Figure \ref{fig:slit} for explanation of regions \\
	b. Einstein A-coefficients from \citet{1998ApJS..115..293W}. \\
	c. Line width values were calculated from inspection of each line profile individually \\ and may differ from values given by the fits in Figure \ref{fig:lines}
\end{table*}


%

Figure \ref{fig:lines} shows the spectra for the S(1) and S(5) lines. We measured flux values for the S(1) and S(5) lines in the \enquote{ridge} and \enquote{shady} regions, but we detected no emission in the \enquote{sunny} region (Table \ref{tab:results}). We calculated the flux values shown in Table \ref{tab:results} by calculating the area of the Gaussian fits of the line spectra shown in Figure \ref{fig:lines}.The line widths used to calculate these areas are shown in Table \ref{tab:results}. We compared this with the direct line integration method and found agreeing flux values. The flux uncertainties are derived from the uncertainties of the fitted Gaussian parameters. The integrated intensity of the S(1) line in the \enquote{shady} region is a factor of 2 larger than that in the \enquote{ridge} region which indicates warmer gas in the \enquote{ridge} region. The integrated intensity of the S(5) line is approximately the same in both regions. The S(1) integrated intensity in the \enquote{ridge} region agrees with the S(1) ISO SWS intensity from \citet{2009MNRAS.400..622T} while the ISO SWS S(5) value is approximately twice the intensity of the EXES S(5) intensity values in both regions.

\subsection{Rotation diagram}
\begin{figure*}
	\centering
    \includegraphics[scale=0.36,angle=0]{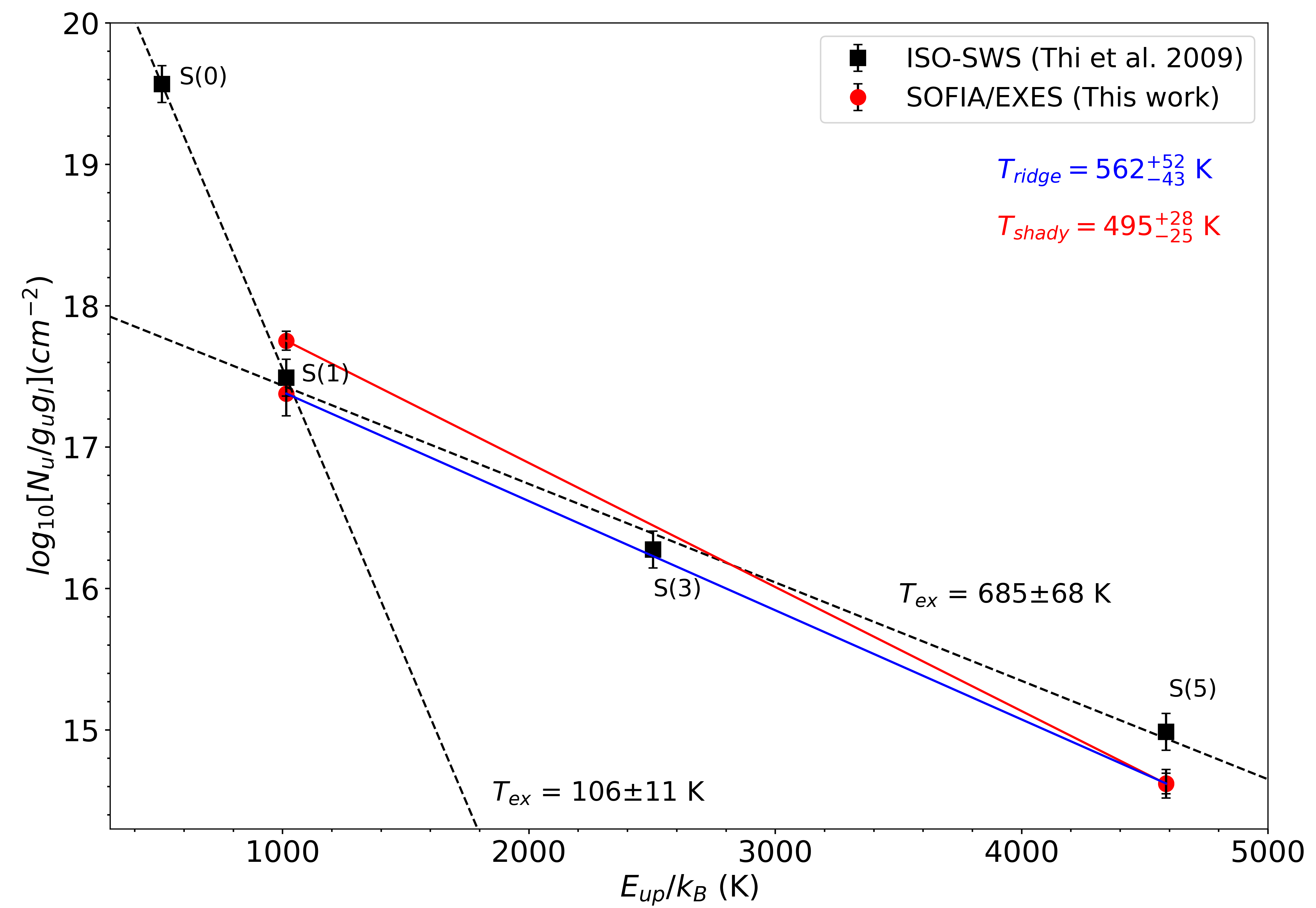}
	\caption{Rotational diagram created with ISO-SWS (solid black rectangles) and EXES data (solid red circles). The two temperatures seen by ISO/SWS data \citep{2009MNRAS.400..622T} are labeled and the temperatures of ridge and shady regions estimated with EXES data also shown with the text in red and blue colors.}\label{fig:excit_diag}
\end{figure*}

We calculated the population of the upper level, $N_{u}$, using the equation:

\begin{equation}
I_{ul} = N_u A_{ul} h \nu_{ul} / 4\pi \;,
\label{eq:cd}
\end{equation}

where $I_{ul}$ is the total intensity, $A_{ul}$ is the Einstein A-coefficient, and $\nu_{ul}$ the frequency of the transitions from the upper to lower levels. We constructed an $\rm H_{2}$ excitation diagram (see Figure \ref{fig:excit_diag}) by plotting the log of the population of the upper level divided by the upper level degeneracy and nuclear spin degeneracy ($g_{u}$, $g_{I}$ respectively) versus the temperatures of the upper energy levels which are 1015 K for S(1) and 4586 K for S(5). The excitation temperature at local thermodynamic equilibrium, $T_{ex}$, can be related to the population of the upper level, $N_{u}$, by the equation:
\begin{equation}
N_{u} = N(H_2) g_I g_u e^{-E_{ul} / T_{ex}} / Q(T_{ex}) \;,
\label{eq:temp}
\end{equation}
where $E_{ul}$ is the energy of the transition expressed in Kelvin, N($H_{2}$) is the $\rm H_{2}$ column density and Q($T_{ex}$) is the partition function. 
In the rotational excitation diagram shown in Figure \ref{fig:excit_diag}, we overplottetd EXES data (red solid circles) with the data taken from \citet{2009MNRAS.400..622T} (black filled squares). 

Based on these two lines, we derive excitation temperatures of T$_{ridge}$=562$^{+52}_{-43}$K and T$_{shade}$=495$^{+28}_{-25}$K. The errors given here are 1$\sigma$ level. This shows that the ridge of IC\,63 seen with EXES observations is at a similar temperature (within uncertainties) as the high temperature reported by \citet{2009MNRAS.400..622T} (T$_{ex}$=685$\pm$68K). However, our results shows that the shady region of IC\,63 is somewhat cooler than their reported temperature. The inferred $\rm H_{2}$ column density from the excitation temperatures of the ridge and shady regions are $\rm 9.8\times 10^{19}~cm^{-2}$ and $\rm 3.6\times 10^{20}~cm^{-2}$, respectively. We could not derive excitation temperature in the sunny side of the cloud as we did not detect either the S(1) or S(5) lines there (see Figure \ref{fig:lines}).

As shown in left panel of Figure \ref{fig:slit}, the SWS footprint is significantly larger than the EXES slit width and located further south - and therefore closer the to illuminating star $\gamma$\ Cas. Because of the area of the SWS aperture, it encompasses gas both on the near and far side of the fluorescent ridge - as seen from $\gamma$\ Cas.  Based on the location and size differences we hypothesize that the warm gas probed by the ISO/SWS measurement may possibly represent higher illumination regions, or a more \enquote{sunny} location and that the EXES observation probe gas with a higher extinction toward the star or a more \enquote{shady} spot if the actual distance of ISO slit is lesser than that of EXES slit from $\gamma$\ Cas. However, the projected distances from $\gamma$\ Cas of the ISO and EXES slits are almost the same. The cold temperature (T$_{ex}$=106$\pm$11\,K) seen in the ISO data would then be tracing well-shielded gas behind a dense clump. This hypothesis can be tested by extending the wavelength and spatial coverage to probe both the $\rm H_{2}$ S(1)/S(0) temperature in the EXES location and by isolating more spatial locations on either side of the PDR ridge.  We are currently pursuing such observations.

\begin{figure}
	\centering
    \includegraphics[scale=0.5,angle=0]{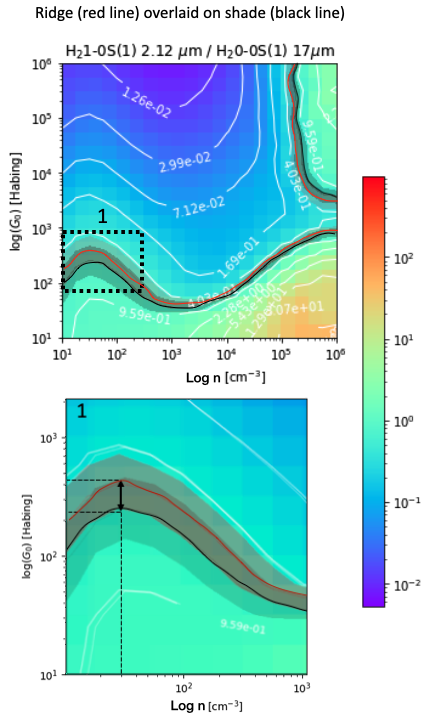}
	\caption{\textbf{Upper} panel shows the contour plot of $\rm H_2$ 1-0 S(1)/0-0 S(1) line ratios in the ridge (red line) and shade (black line) regions as functions of density and UV intensity $\rm G_0$ with respect to the Habing field, adapted from \citet{kaufman2006}. The width of the shade represents $\rm 3\sigma$ confidence. Dashed box 1 shown in the in the upper panel is zoomed-in in the lower panel.}\label{fig:model}
\end{figure}

\subsection{PDR modeling}
We used the PDR toolbox \citep{kaufman2006, pound2008} to model our observed line ratio $\rm H_{2}$ 1-0 S(1)/0-0 S(1). This model takes into account both thermal and UV excitation for the $\rm H_{2}$ level population. We have plotted the predicted values for different values of FUV radiation field intensity ($\rm G_{0}$) and volume density (n) in Figure \ref{fig:model}, where we have also over-plotted the observed line ratios. The upper panel of Figure \ref{fig:model} shows the observed values over-plotted for the ridge and shade sides using red and black lines, respectively. The shaded regions show 3$\sigma$ uncertainties limits in the mean values. We find values of $\rm G_{0}$, in shade and ridge of $\sim 200$ and $\sim 400$ in Habing units, respectively. Using conversion from Habing units to Draine unit ($\rm Draine = 1.7\times Habing$) \citep{Wolfire2011}, these values are 340 and 680 Draine units in shade and ridge, respectively. The value of $\rm G_{0}$ in ridge is consistent to the $\rm G_{0}$ value of 650 Draine units reported by \citet{jansen1994} in IC\,63.

The dashed box labeled as 1 in the upper panel of Figure \ref{fig:model} clearly show different $\rm G_{0}$ values for the ridge and shade sides (red and black solid lines). This region is zoomed-in in the lower panel of he figure where the $\rm G_{0}$ gap is indicated by double headed arrows. We used this difference in $\rm G_{0}$ to estimate the gas density in the PDR by adapting the radiative transfer equation:

\begin{equation}
    {G_{0}}_{shade} = e^{-\tau}{G_{0}}_{ridge}
\end{equation}
where ${G_{0}}_{shade}$ and ${G_{0}}_{ridge}$ are the model returned values of $\rm G_{0}$ for the line-ratios observed in the shade and ridge sides of IC\,63 PDR, respectively. $\tau$ is optical depth of the region for the wavelength determining the line ratio (i.e. gas temperature). Using $\rm G_{0}$ values in eq. 3, we estimated an optical depth of 0.4.

In PDRs, when FUV photons are absorbed, the grain may be ionized (photoelectric effect) and part of the photon energy is carried away by the electron, heating the gas. Assuming a work function of 8 eV \citep{weingartner&jordan2008} corresponds to a wavelength of $\sim$\SI{1550}{\angstrom} we can  estimated the gas density in the PDR using the optical depth estimated above.  Using relation $\rm A_{\SI{1550}{\angstrom}}/A_{v}$ = 2.64 from \citet{whittet2003}, and $\rm A_{\SI{1550}{\angstrom}} = 1.086\times \tau$, we estimated $\rm A_{v}$ of $\sim$ 0.16. This value of $\rm A_{v}$ and total-to-selective extinction ($\rm R_{v}=3.1$) together with the relationship $\rm N(H) = 1.87\times 10^{21}A_{v}$ and $\rm N(H_{2}) = 0.5 N(H)$ from \citet{bohlin1978}, yields a column density $\rm N(H_{2})$ of $\rm 2.9\times 10^{20} cm^{-2}$. Assuming a constant space density and that the absorption takes place over a region of projected length L we can estimate the space density. If we assume L to be the EXES slit size of the $H_{2}$ S(1) observations i.e. 19$\arcsec$ shown in Figure \ref{fig:slit}, we estimate a volume density $\rm n(H_{2})$ of $\rm 5\times 10^{3} cm^{-3}$. 

We also estimated column densities in shade ($\rm 3.6\times 10^{20}~cm^{-2}$) and ridge ($\rm 9.8\times 10^{19}~cm^{-2}$) regions of IC\,63 from the rotation diagram (Sec. 3.1). Using these values over a projected length L at the location of EXES slit, we derived space densities of $\rm 6.2\times 10^{3}~cm^{-3}$ and $\rm 1.7\times 10^{3}~cm^{-3}$ in shade and ridge sides, respectively. These density values are consistent with our findings when assuming heating due to the photoelectric effect. 
\citet{jansen1994} presented far-infrared spectroscopic measurements of IC\,63. They measured line ratios of CO, $\rm HCO^{+}$, HCN, CS and HCHO suggesting that cloud is warm with a temperature $\approx$50\,K, and the density of the gas is $\rm 5\times 10^{4}~cm^{-3}$. This value is ten times higher than our estimation above. It might be possible that the values reported in \citet{jansen1994} are biased towards denser clumps. 

As it can be noticed in Figure \ref{fig:model}, the modeled $\rm H_{2}$ 1-0 S(1)/0-0 S(1) line ratio gives a completely different answer for the density. The  density value obtained here is of the order of $\rm 2\times 10^{2}~cm^{-3}$. It is also noticed by \citet{2009MNRAS.400..622T} that the $\rm H_{2}$ S(3)/S(1) line ratio gives a completely different answer for FUV radiation strength and density. \citet{2009MNRAS.400..622T} uses the explanation given by \citet{bertoldi1997} and \citet{allers2005} to counter these discrepancies. It is suggested that either the $\rm H_{2}$ formation rate on grains needs to be larger or the photoelectric heating efficiency needs to be increased at high temperatures, shifting the transition zone of H $\rightarrow$ H$_2$ closer to the warm edge of PDRs. High photoelectric heating efficiencies have been calculated by \citet{weingartner1999} based on an enhanced dust-to-gas ratio in the PDR due to gas–grain drift. We agree with the findings and possible causes explained by \citet{2009MNRAS.400..622T}. The photoelectric heating might be underestimated at hotter regions which may cause in discrepancies in the density estimation. Also, we did not include a detailed analysis on $\rm H_{2}$ formation rate in this work because that is beyond the purpose of this paper. An underestimation in assuming the rate might also cause the discrepancies between model and observed densities.

\section{Discussion}\label{sec:diss}
Our results show that SOFIA/EXES is capable of investigating the spatial variation in temperatures in such regions. Although we do not see a big difference from temperatures values of \citet{2009MNRAS.400..622T}, we are capable of better resolving the spatial locations in the IC\,63 PDR. The spatial coverage of ISO data used by \citet{2009MNRAS.400..622T} was $\approx 25^{\prime\prime}$ or 0.025\,pc at the distance of IC\,63.

\citet{andrews2018} investigated the physical conditions of UV-illuminated surface of IC\,63 using $Spitzer$ archival data of molecular hydrogen. They used the short and long wavelength (SL, 5.2–14.5 \micron; LL, 14.0–38.0 \micron) low resolution modules (R $\sim$ 60-130). The spatial coverage of SL mode was $\rm \approx0.8\times1.0~arcmin^{2}$ (i.e., $\rm 0.046\times 0.058\,pc^{2}$ at the distance of IC\,63) and the spatial coverage of LL mode on the tip of IC\,63 nebula was $\rm \approx2.8\times2.4~arcmin^{2}$ (i.e., $\rm 0.16\times0.14\,pc^{2}$).

They found the cooler temperature (T=207$\pm$30\,K) associated to higher column density of $\rm 2.3\times10^{20}\,cm^{-2}$ and the warmer component (T=740$\pm$47\,K) is associated to lower density $\rm 9.7\times10^{17}\,cm^{-2}$ region. Their values are in good agreement with the hot component (i.e., T=685 $\pm$68\,K) of \citet{2009MNRAS.400..622T}. But these values are higher than what we found as hot component (i.e., T$_{ridge}$=562$^{+52}_{-43}$K) with SOFIA/EXES higher spatial resolution observations. This suggests that \citet{2009MNRAS.400..622T} and \citet{andrews2018} get an averaged value of temperature in the bigger aperture of the ISO/SWS and $Spitzer$/IRS, respectively. Whereas EXES observations are further capable of resolving temperature of PDRs.

\citet{soam21a} showed that the collisional disalignment rate of the dust grains causing the observed polarization follow a bifurcated relation with respect to the gas density. The two observed sequences in derived disalignment rate correspond to lines of sight in front and behind gas clumps, as seen in the HCO$^+$(J=1-0) map of the cloud.  A two temperature structure in the gas, caused by variable internal shading by molecular clumps, would explain this observed bifurcation.

We used PDR modeling in Section 3.2 to estimate the density of the gas adapting the radiative transfer equation and assuming that the heating is due to the photo-electric effect.  The density of gas is found to be $\rm 5\times 10^{3} cm^{-3}$. Our estimation of density is consistent to the value reported by \citet{jansen1994}. \citet{habart2011} in their investigation on number of PDRs stated that for the physical conditions prevailing in our PDRs, the heating is mainly due to the photoelectric effect. Authors also report that for a range of $\rm H_2$ formation rates and densities in various PDRs, the gas temperature vs. PDR depth profile remains insensitive to the fraction of molecular hydrogen. However, a higher formation rate of molecular hydrogen may shift the photodissociation front towards higher temperatures.

The non-detection of H$_2$ pure-rotational emission in the \enquote{sunny} part of the nebula is, in retrospect, not surprising.  The locations of the \enquote{sunny} and \enquote{shade} regions were chosen based on mid/far infrared contiunuum and and H$_2$ v=1-0 S(1) line imaging.  To allow for the presence of H$_2$ not detected by the fluorescent emission - possibly as a result of a highly clumpy medium on small scales -  (\enquote{sunny}) locations star-ward of the fluorescent ridge were included in our observations.  However, given the sharp transition expected from atomic to molecular form of hydrogen, caused by the self-shielding of the molecule \citep[e.g.][]{federman1979} the non-detection of H$_2$ in the \enquote{sunny} regions is consistent with this sharp transition and indicates that the PDR is not very clumpy. 

\raggedbottom
\section{conclusions}\label{sec:con}
We have re-investigated the temperature in the IC\,63 PDR using pure-rotational molecular hydrogen observations of S(1) and S(5) lines using SOFIA/EXES observations. A similar investigation was done by \citet{2009MNRAS.400..622T} in this nebulae using lower spatial resolution ISO/SWS observations. The higher spatial resolution of EXES over SWS enables us to spatially resolve temperature in this PDR. \citet{2009MNRAS.400..622T} reported two components of gas using S(0), S(1), S(3), and S(5) pure rotational lines of molecular hydrogen. The warm component is found to be at T$_{ex}$=106$\pm$11\,K and hot gas component is seen at T$_{ex}$=685$\pm$68\,K. We divided IC\,63 PDR into \enquote{shade}, \enquote{ridge} and \enquote{sunny} sides for our investigation. By constructing rotation diagram using S(1) and S(5) lines data from EXES, we obtained temperature of T$_{ex}$=$562^{+52}_{-43}$\,K towards the \enquote{ridge} and T$_{ex}$=$495^{+28}_{-25}$\,K in \enquote{shady} side. The PDR toolbox code was used to model the line ratios of detected emission in $\rm H_{2}$ 1-0 S(1) and 0-0 S(1) transitions. Our model suggests a lower value of FUV radiation ($\rm G_{0}$) in the shade as compared the one in ridge. We used this damping in FUV radiation to estimate the optical depth of the PDR. Our results emphasize that SOFIA/EXES has the capability to resolve and provide detailed information on gas temperature and density of such regions. We are attempting to expand this work through observations of the S(2) and S(4) transitions from IC\,63.

\bigskip
\bigskip
We thank the anonymous referee for helping us to improve the content of the paper considerably. Authors also thank EXES/SOFIA team for the data. SOFIA is jointly operated by the Universities Space Research Association, Inc. (USRA), under NASA contract NNA17BF53C, and the Deutsches SOFIA Institut (DSI) under DLR contract 50 OK 0901 to the University of Stuttgart. EXES is supported by NASA agreement 80NSSC19K1701. Financial support for the work in this paper was provided by NASA through award $07_{-}$0040 issued by USRA, and by the NSF through Grant-1715876. J.K. acknowledges funding from the Moses Holden Scholarship in support of his PhD. A.S. thanks Mark Wolfire and Marc Pound for adding required line ratio to their PDR tool box. We thank Doug Hoffman (USRA/NASA Ames) for helping in calibration of the data. 

\facility{EXES/SOFIA, CFHT}

\software{Astropy \citep{2013A&A...558A..33A}, Matplotlib \citep{Hunter:2007}, SciPy \citep{2020SciPy-NMeth}, NumPy \citep{2020NumPy-Array}}


\bibliography{ref_H2}{}
\bibliographystyle{aasjournal}
\end{document}